%% file: DDPM_newformat_v2.3.tex
\documentclass[12pt,letterpaper,onecolumn]{article}
\usepackage{cite,graphicx,amssymb,color,epsfig,amsmath,times,bm,caption}
\usepackage{subfig}
\usepackage{makecell}
\usepackage{amssymb}
\usepackage{algorithm,algorithmic}
\newcommand{\bb}{\bm}
\usepackage[sort&compress, numbers]{natbib}

\input{defs}

\begin{document}
\author{Kuang Gong$^{1,2}$,  Keith A.  Johnson$^{1}$,  Georges El Fakhri$^{1}$, Quanzheng Li$^{1,2}$,  and Tinsu Pan$^3$}
\title{PET image denoising based on denoising diffusion probabilistic models}
\maketitle

\section*{Abstract}
\vspace{1cm}
\textbf{Purpose:} Due to various physical degradation factors and limited counts received, PET image quality needs further improvements. The denoising diffusion probabilistic models (DDPM) are distribution learning-based models, which try to transform a normal distribution into a specific data distribution based on iterative refinements. In this work, we proposed and evaluated different DDPM-based methods for PET image denoising.

\noindent
\textbf{Methods:} Under the DDPM framework, one way to perform PET image denoising is to provide the PET image and/or the prior image as the input.  Another way is to supply the prior image as the network input with the PET image included in the refinement steps, which can fit for scenarios of different noise levels. 120 $^{18}$F-FDG datasets and 140 $^{18}$F-MK-6240 datasets were utilized to evaluate the proposed DDPM-based methods.

\noindent
\textbf{Results:} Quantification shows that the DDPM-based frameworks with PET information included can generate better results than the nonlocal mean and Unet-based denoising methods.  Adding additional MR prior in the model can help achieve better performance and further reduce the uncertainty during image denoising. Solely relying on MR prior while ignoring the PET information can result in large bias. Regional and surface quantification shows that employing MR prior as the network input while embedding PET image as a data-consistency constraint during inference can achieve the best performance.

\noindent
\textbf{Conclusions:} DDPM-based PET image denoising is a flexible framework, which can efficiently utilize prior information and achieve better performance than the nonlocal mean and Unet-based denoising methods.

\noindent {\bf Keywords:} PET image denoising,  Denoising diffusion probabilistic models,   Low-dose PET,  Generative models

\pagebreak

\section*{Introduction}
\label{sec:intro}  
Positron Emission Tomography (PET) is widely used in oncology, cardiology and neurology studies because of its high sensitivity and quantitative merits.  Due to various physical degradation factors and limited counts received,  the signal-to-noise ratio (SNR) and resolution of PET is low,  which comprises its clinical values in diagnosis, prognosis, staging and treatment monitoring.  Additionally,  to improve the hospital's throughput or reduce the radiation exposures to patients,  faster or low-dose PET imaging is desirable,  where the counts received during the scan is even less.  This further challenges our ability to attain high-quality PET images from limited counts.

\noindent For the past decades, various approaches have been explored for PET image denoising,  e.g.,  wavelet \cite{lin2001improving}, highly constrained backprojection (HYPR) \cite{christian2010dynamic},  nonlocal mean \cite{chan2014postreconstruction,dutta2013non},  guided image filtering \cite{yan2015mri}, block matching \cite{ote2020kinetics},  and dictionary learning \cite{wang2016semisupervised}.  These methods are mainly based on specifically designed similarity calculation or small-scale feature extraction/learning.  With the availability of enormous training data and strong computational power,  deep learning-based image processing has been extensively studied recently.  More specifically,  convolutional neural network (CNN)-based PET image denoising has achieved superior performance than other state-of-the-art methods \cite{xiang2017deep,kaplan2019full,cui2019pet,chen2019ultra, hashimoto2019dynamic,da2020micro,schramm2021approximating,mehranian2022image,daveau2022deep}.  However,  over-smoothness of the network output is one pitfall of most CNN-based approaches.  Conditional generative adversarial networks (cGANs)-based PET image denoising have been developed to address this issue \cite{ouyang2019ultra,lei2019whole,zhou2020supervised,song2020pet,sanaat2021deep,xue2022cross},  where the learned discriminative loss can further push the image appearance closer to the ground truth.  One issue of the CNN and cGAN-based methods is that if there are data-distribution mismatches between training and testing data,  the network performance can be compromised.  Transfer learning has been investigated as one approach to address the mismatch issue and has been demonstrated effective in various PET studies \cite{gong2018pet,liu2020noise,chen2020generalization,cui2021populational,zhou2022federated}.   

\begin{figure}[htp]
\centering
\subfloat{\includegraphics[trim=1cm 0.2cm 1cm 0.3cm, clip=true, width=7in]{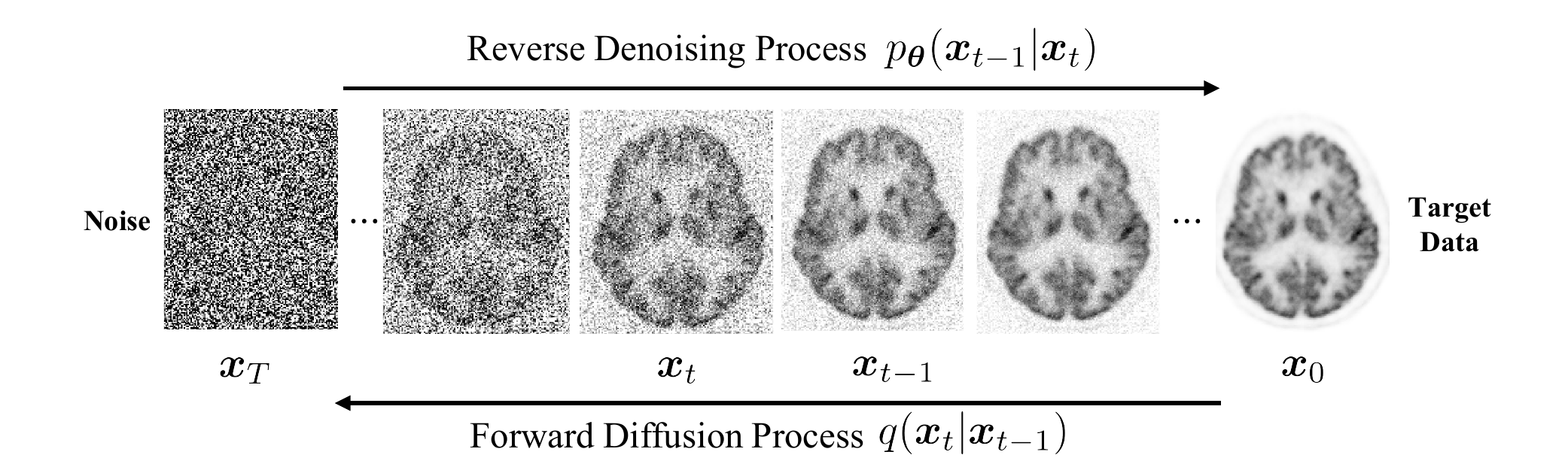}} \\
\caption{ Diagram of the DDPM framework.}
\label{fig:ddpm}
\end{figure}

\noindent The deep learning approaches described above are deterministic models,  where only one denoised image can be obtained based on the pre-specified network input.  The denoising diffusion probabilistic model (DDPM) and score function-based model \citep{ho2020denoising,song2019generative,song2020score},  denoted as diffusion models here, are a category of distribution learning-based models,  which try to transform a normal distribution into a specific data distribution based on iterative refinements.  During the forward pass,  noise was injected in the image gradually.   During the reverse pass,  data samples were generated by removing noise from the image after multiple refinement steps.  Fig.~\ref{fig:ddpm} shows the diagram of the forward and reverse pass.  Additional prior image can be added to the model to conduct conditional image generation.  By running the reverse pass multiple times,   an uncertainty map can also be calculated from the generated data samples,  which can be used to further guide diagnosis and treatment monitoring.  The diffusion models have shown encouraging results in various computer-vision tasks,  e.g.,  image generation \cite{song2019generative,song2020score,ho2020denoising,dhariwal2021diffusion,rombach2022high},   super resolution \cite{saharia2021image,dhariwal2021diffusion},  image in-painting \cite{lugmayr2022repaint},  and MR image reconstruction \cite{jalal2021robust,chung2022score}.

\noindent As the dynamic range of different organs' uptake is high and the noise levels of different scans vary,  the data distribution of PET images are more complicated than natural images.  Additionally,   apart from PET image itself,  there are patients' anatomical prior information (e.g., CT or MR images) available from the same or separate scans.  Thus,   further evaluating the diffusion models and exploring approaches to embed the patient's prior information for PET image denoising deserve further investigations. In this work,  we evaluated DDPM for PET image denoising and further explored  different approaches of embedding the PET and the prior MR information in DDPM.  Nonlocal mean and  Unet-based PET image denoising methods were employed as the reference methods.  120 brain $^{18}$F-FDG datasets and 140 $^{18}$F-MK-6240 datasets were utilized in two evaluation studies to evaluate the performance of different methods based on global and regional quantitative metrics.



\section*{Materials and methods}
\noindent \textbf{DDPM model}

\noindent Suppose the target data $\bb{x}_0\sim q(\bb{x}_0)$.  During the forward process,  we can define a Markov diffusion process $q$ that adds Gaussian noise to $\bb{x}_0$ in each step,  which can be written as \cite{ho2020denoising}
\begin{equation}
q(\bb{x}_1, .., \bb{x}_T|x_0) = \prod_{t=1}^Tq(\bb{x}_t|\bb{x}_{t-1}),  \qquad q(\bb{x}_t|\bb{x}_{t-1}) =  \mathcal{N}(\bb{x}_t;\sqrt{1-\beta_t}\bb{x}_{t-1}, \beta_t \bb{I}).
\end{equation}
One property of the forward process $q$ is that it allows us to sample at any arbitrary step directly conditioned on $\bb{x}_0$.  Denoting $\alpha_t = 1-\beta_t$ and $\bar{\alpha}_t = \prod_{l=1}^t\alpha_l$, we can have 
\begin{equation}
q(\bb{x}_t|\bb{x}_{0}) =  \mathcal{N}(\bb{x}_t;\sqrt{\bar{\alpha}_t }\bb{x}_{0}, (1-\bar{\alpha}_t)\bb{I}),   \qquad \bb{x}_t = \sqrt{\bar{\alpha}_t }\bb{x}_{0} + (1-\bar{\alpha}_t)\bb{\epsilon},
\label{eps}
\end{equation}
where $\bb{\epsilon} \sim  \mathcal{N}(\bb{0},\bb{I})$ .  Based on Bayes theorem, the posterior $q(\bb{x}_{t-1}|\bb{x}_t,\bb{x}_0)$ also follows Gaussian distribution,  
\begin{equation}
q(\bb{x}_{t-1}|\bb{x}_t, \bb{x}_0)= \mathcal{N}(\bb{x}_{t-1};\tilde{\bb{\mu}}_t(\bb{x}_t,\bb{x}_0),  \frac{1-\bar{\alpha}_{t-1}}{1-\bar{\alpha}_{t}}\beta_t\bb{I}), 
\end{equation}
\begin{equation}
\text{with} \qquad \tilde{\bb{\mu}}_t(\bb{x}_t,\bb{x}_0) = \frac{\sqrt{\bar{\alpha}_{t-1}}\beta_t}{1-\bar{\alpha}_t}\bb{x}_0 +  \frac{\sqrt{\alpha_t}(1-\bar{\alpha}_{t-1})}{1-\bar{\alpha}_t}\bb{x}_t.
\label{mu}
\end{equation}
If we want to sample from the target data distribution $q(\bb{x}_0)$,  we can first sample from $q(\bb{x}_T)$,  which is an isotropic Gaussian distribution given a large enough $T$.  Then based on the posterior distribution $q(\bb{x}_{t-1}|\bb{x}_t)$,  we can obtain samples of $\bb{x}_0$.  However,  $q(\bb{x}_{t-1}|\bb{x}_t)$ is not computable as the distribution of $\bb{x}_0$ is unknown.  The DDPM framework tries to approximate $q(\bb{x}_{t-1}|\bb{x}_t)$ by $p_{\bb{\theta}}(\bb{x}_{t-1}|\bb{x}_t)$ through a network with parameter $\bb{\theta}$,  where
\begin{equation}
p_{\bb{\theta}}(\bb{x}_{t-1}|\bb{x}_{t}) =  \mathcal{N}(\bb{x}_{t-1};\tilde{\bb{\mu}}_{\bb{\theta}}(\bb{x}_t, t), 
\sigma^2_t\bb{I}).
\label{eq:p_theta}
\end{equation}
Instead of directly approximating $\tilde{\bb{\mu}}_{\theta}(\bb{x}_t, t)$ by a neural network,  Ho {\it{et al}} \cite{ho2020denoising} proposed to approximate the noise $\bb{\epsilon}$ in equation~(\ref{eps}) by a network as $\bb{\epsilon}_{\bb{\theta}}(\bb{x}_t,t)$,  which can also be interpreted as the gradient of the data log-likelihood,  known as the score function.  Based on equations~(\ref{eps}) and (\ref{mu}),  $\tilde{\bb{\mu}}_{\bb{\theta}}(\bb{x}_t, t)$ can expressed as 
\begin{equation}
\tilde{\bb{\mu}}_{\bb{\theta}}(\bb{x}_t,t) = \frac{1}{\sqrt{\alpha_t}}\left[ \bb{x}_t- \frac{\beta_t}{\sqrt{1-\bar{\alpha}_t}}\bb{\epsilon}_{\bb{\theta}}(\bb{x}_t,t)\right].
\label{eq:new_mu}
\end{equation}
Based on the trained score function $\bb{\epsilon}_{\hat{\bb{\theta}}}(\bb{x}_t,t)$, and equations~(\ref{eq:p_theta}) and (\ref{eq:new_mu}), each refinement step during inference is
\begin{equation}
\bb{x}_{t-1} = \frac{1}{\sqrt{\alpha_t}}\left[ \bb{x}_t- \frac{\beta_t}{\sqrt{1-\bar{\alpha}_t}}\bb{\epsilon}_{\hat{\bb{\theta}}}(\bb{x}_t,t)\right] +\sigma_t \bb{z}, \quad \text{where} \quad \bb{z} \sim  \mathcal{N}(\bb{0},\bb{I}).
\label{eq:refine}
\end{equation}

\noindent \textbf{Conditional PET image denoising based on DDPM}

\noindent The original DDPM stated above is for unconditional image generation.  For PET image denoising, the noisy PET image $\bb{x}_{\text{noisy}}$ exists and the patient's prior image $\bb{x}_{\text{prior}}$ might also be available.  We can directly supply $\bb{x}_{\text{noisy}}$  and $\bb{x}_{\text{prior}}$ (if available) as the additional network input of the score function,  as was done in image super-resolution \cite{saharia2021image}, and the score function $\bb{\epsilon}_{{\bb{\theta}}}(\bb{x}_t,t)$ changes to $\bb{\epsilon}_{{\bb{\theta}}}(\bb{x}_t,t, \bb{x}_{\text{noisy}}, \bb{x}_{\text{prior}})$. This framework requires specific low- and high-quality training pairs during the score function training.  Another approach is to only supply $\bb{x}_{\text{prior}}$ as the additional input to the score function,  while including $\bb{x}_{\text{noisy}}$ during inference.  Compared to the previous approach,  this approach does not need low- and high-quality pairs during training and can thus work for noisy PET images of various noise levels during inference.  In this approach, we can approximate $q(\bb{x}_{t-1}|\bb{x}_t)$  by $p_{\bb{\theta}}(\bb{x}_{t-1}|\bb{x}_t, \bb{x}_{\text{noisy}}, \bb{x}_{\text{prior}})$ instead of $p_{\bb{\theta}}(\bb{x}_{t-1}|\bb{x}_t)$ in the original DDPM framework, which can be rewritten as 
\begin{equation}
p_{\bb{\theta}}(\bb{x}_{t-1}|\bb{x}_t, {\bb{x}_{\text{noisy}}},\bb{x}_{\text{prior}}) \propto p_{\bb{\theta}}(\bb{x}_{t-1}|\bb{x}_t, \bb{x}_{\text{prior}}) p({\bb{x}_{\text{noisy}}}| \bb{x}_{t-1}, \bb{x}_{t},\bb{x}_{\text{prior}}).
\end{equation}
Suppose the PET image noise follows Gaussian distribution and is independent of $\bb{x}_{\text{prior}}$,   we can have 
\begin{equation}
p({\bb{x}_{\text{noisy}}}| \bb{x}_{t-1}, \bb{x}_{t},\bb{x}_{\text{prior}})\approx \mathcal{N}(\bb{x}_{t},\sigma^2_d\bb{I}),
\label{eq:data_constraint}
\end{equation}
where $ \sigma_d$ indicates the noise level of ${\bb{x}_{\text{noisy}}}$.  As for $ p_{\bb{\theta}}(\bb{x}_{t-1}|\bb{x}_t, \bb{x}_{\text{prior}})$,  similar to equations~(\ref{eq:p_theta}) and (\ref{eq:new_mu}), $\bb{\epsilon}$ was approximated by $\bb{\epsilon}_{{\bb{\theta}}}(\bb{x}_t,t, \bb{x}_{\text{prior}})$, where $\bb{x}_{\text{prior}}$ was supplied as an additional input to the score function.   Based on equation (10) from Ref. \citenum{dhariwal2021diffusion} and equation~(\ref{eq:data_constraint}), the refinement step during inference becomes
\begin{equation}
\bb{x}_{t-1} = \frac{1}{\sqrt{\alpha_t}}\left[ \bb{x}_t- \frac{\beta_t}{\sqrt{1-\bar{\alpha}_t}}\bb{\epsilon}_{\hat{\bb{\theta}}}(\bb{x}_t,t,\bb{x}_{\text{prior}})\right] - \frac{{\sigma}^2_t}{\sigma^2_d }(\bb{x}_{\text{noisy}} - \bb{x}_t) +\sigma_t \bb{z}.
\label{eq:refine_v2}
\end{equation}

\noindent \textbf{Datasets}

\noindent  Two experiments based on $^{18}$F-FDG and $^{18}$F-MK-6240 datasets were conducted to evaluate the performance of DDPM for PET image denoising and  also explore different approaches of embedding prior information.  Firstly, 120 brain $^{18}$F-FDG datasets acquired from the GE DMI PET-CT scanner were utilized.  85 datasets were used for training,  5 datasets for validation and the remaining 30 datasets for testing.  Normal-dose datasets were used as the ground truth and low-dose datasets were generated by extracting 1/4 events from the normal-dose listmode datasets.  All the datasets were reconstructed into a matrix of $256 \times 256 \times 89$ with a voxel size of $1.17 \times 1.17 \times 2.79 $ mm${^3}$ based on the ordered subset expectation maximization (OSEM) algorithm (3 iterations and 17 subsets) including point spread function (PSF) and time of flight (TOF) modeling.  The images were later scaled to the standardized uptake value (SUV) unit before being further processed using different methods.  For the proposed DDPM-based method,  low-dose PET images were supplied as the network input and this method is denoted as DDPM-PET.

\noindent The other experiment is based on $^{18}$F-MK-6240 datasets with corresponding T1-weighted MR images.  In total 140 dynamic $^{18}$F-MK-6240 datasets acquired on the GE DMI PET/CT scanner were utilized in this experiment.  For each dataset,  the  events from 0 min to 10 min post injection, were extracted and combined to construct the normal-dose dataset.  These early-time-frame datasets mainly contain brain perfusion information,  which is actively investigated as a potential biomarker for neurodegenerative diseases \cite{tiepolt2016early,hammes2017multimodal,visser2020tau}.  The extracted events were further down sampled to generate 1/4 low-dose datasets.  Of all the 140 data pairs, 116 were employed for training,  4 for validation and the remaining 20 for testing.  The image reconstruction algorithm, matrix size and pixel size is the same as the above-mentioned $^{18}$F-FDG datasets.  Additional T1-weighted MR images were acquired from the Siemens 3T MR scanner and was registered to the PET images through rigid registration by ANTS \cite{avants2009advanced}.  In this experiment,  we have different DDPM-based methods by varying embedding approaches of the low-dose PET and MR prior images.  The method using PET image only as the network input is denoted as DDPM-PET (the same as in the $^{18}$F-FDG experiment).  The method using MR image only as the input is denoted as DDPM-MR.  The method using both PET and MR images as the input is denoted as DDPM-PETMR. Finally, the method using the MR image as the network input while using the PET image in the data consistency item,  as shown in equation~(\ref{eq:refine_v2}), is denoted as DDPM-MR-PETCon.  

\noindent \textbf{Training details and reference methods}

\noindent The network used to generate the score function $\epsilon_{\theta}$ is a Unet structure with attention and residual blocks as described in \cite{dhariwal2021diffusion}.   Two neighboring axial slices were additionally supplied as the network input, to avoid the axial artifacts.  During network training,  the batch size is 20,  the network input is 256 $\times $ 256,   the input channel size is 3 for DDPM-PET,  DDPM-MR and DDPM-MR-PETCon,  and the input channel size is 6 for DDPM-PETMR.  For each DDPM-based method,  the training took around 10 days using 4 NVIDIA RTX 8000 GPUs.  The number of time steps is 1000 and the inference took around 50 mins based on 1 GPU.  

\noindent For both $^{18}$F-FDG and $^{18}$F-MK-6240 experiments,  the nonlocal mean (NLM) and the Unet-based denoising were adopted as the reference methods.  For the Unet, the training datasets and the number of training parameters are the same as the DDPM-based methods.  For the $^{18}$F-FDG experiment,  the similarity in the NLM method was calculated based on the low-dose PET image.  The three-channel  input of the Unet is also the low-dose PET image.  In the $^{18}$F-MK-6240 experiment,  the similarity in the NLM method was calculated based on the MR prior image.  The six-channel  input of the Unet contains both the low-dose PET and MR prior images.

\noindent \textbf{Data analysis}

\noindent For $^{18}$F-FDG  and $^{18}$F-MK-6240 test datasets,  the peak signal-to-noise ratio (PSNR) and structural similarity index measure (SSIM) were calculated for different methods with normal-dose PET images as the ground truth.  The Wilcoxon signed-rank tests of PSNR and SSIM were conducted to compare different methods.  In the $^{18}$F-MK-6240 experiment,  as the patient's MR image is available,   Freesurfer \cite{fischl2012freesurfer} was used for cortical parcellation based on the MR image to obtain the region of interests (ROIs).  Further quantification of the cortical ROIs based on PSNR was performed.  The Braak stage-related cortices---hippocampus,  entorhinal,  parahippocampal,  amygdala,  inferior temporal,  fusiform,  posterior cingulate,  lingual,  precuneus,  insula, pericalcarine,  cuneus and precentral---were chosen as the evaluation ROIs. 

\noindent We have further performed surface analysis to comprehensively evaluate different methods' performance on generating high-quality PET images in the  $^{18}$F-MK-6240 experiment.  Firstly,  the relative PET error, $\text{PET}_{\text{error}} $, was calculated as
\begin{equation}
\text{PET}_{\text{error}} = \frac{\text{PET}_{\text{post}} - \text{PET}_{\text{normal}}}{\text{PET}_{\text{normal}}},
\label{eq:error}
\end{equation}
where $\text{PET}_{\text{post}} $ and $\text{PET}_{\text{normal}}$ are the low-dose image after post-processing and the normal-dose image, respectively.  Afterwards, the surface map of $\text{PET}_{\text{error}} $ was generated for each $^{18}$F-MK-6240 test  dataset and was registered to the FSAverage template in Freesurfer to construct the averaged surface map of all the 20 test datasets.  

\section*{Results}

\begin{figure}[htp]
\centering
\subfloat{\includegraphics[trim=4cm 9.3cm 9.5cm 8.2cm, clip=true, width=6in]{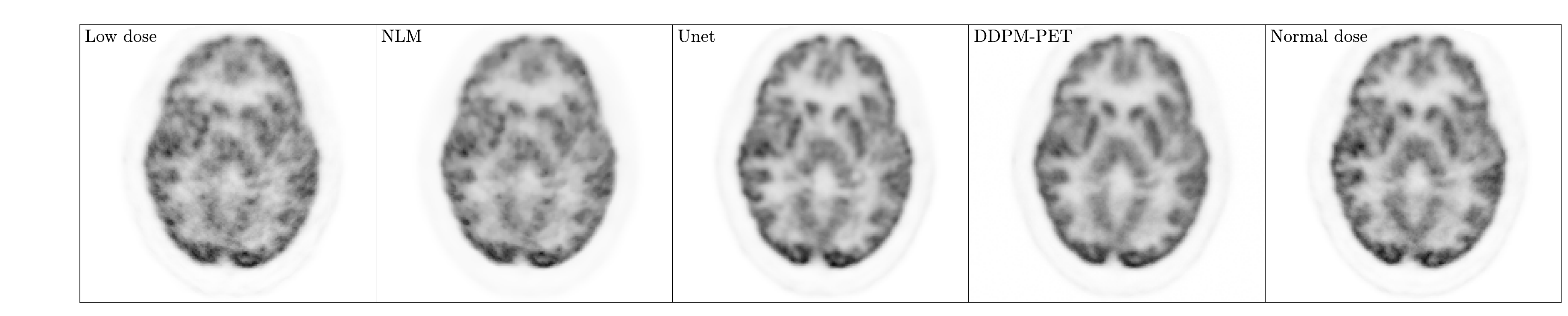}} \\
\subfloat{\includegraphics[trim=4cm 9cm 9.5cm 8.1cm, clip=true, width=6in]{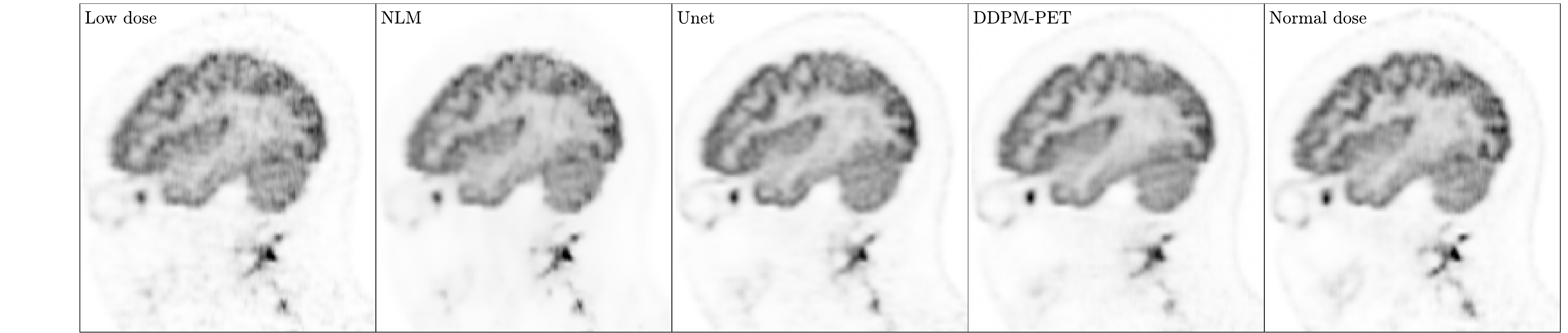}} \\
\subfloat{\includegraphics[trim=4cm 9.2cm 9.5cm 8.2cm, clip=true, width=6in]{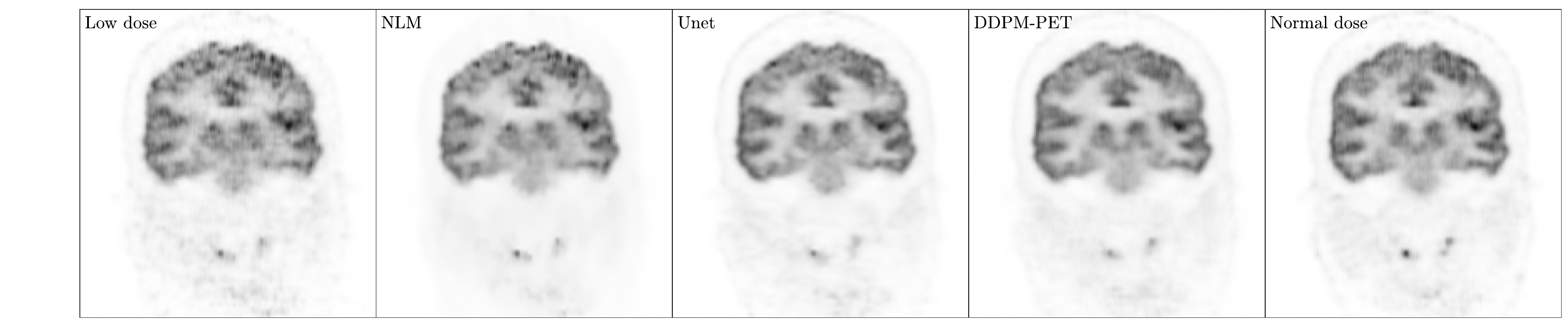}} \\
\caption{Three views of one $^{18}$F-FDG test dataset processed using different methods along with the 1/4 low-dose image (1st column) and the normal-dose image (last column).   }
\label{fig:fdg_image}
\end{figure}

\begin{figure}[htp]
\centering
\subfloat{\includegraphics[trim=0cm 0.8cm 0cm 0cm, clip=true, width=7in]{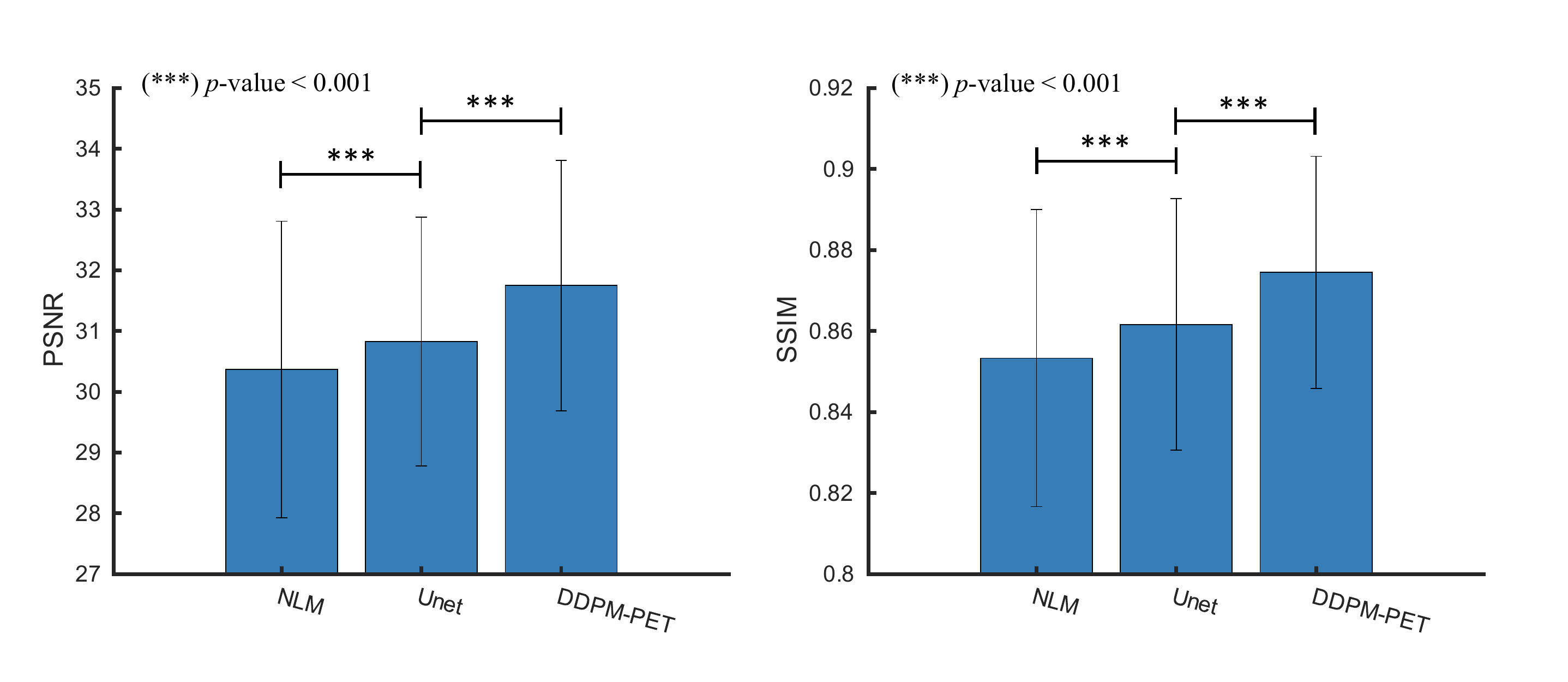}} 
\caption{The PSNR (left) and SSIM (right) values using different methods calculated based on 30 $^{18}$F-FDG test datasets.  *** located at the top of the bar plot represents $p$-value$<0.001$.}
\label{fig:fdg_psnr_ssim}
\end{figure}

Fig. ~\ref{fig:fdg_image} shows three views of one $^{18}$F-FDG dataset processed with different methods. It can be observed that both the Unet and the DDPM-PET methods can generate results with better cortical details and smaller noise in the while matter region compared to the NLM method.  Fig.~\ref{fig:fdg_psnr_ssim} presents the PSNR and SSIM values of different methods based on 30 test datasets, which show that the DDPM-PET method has better quantification results than other reference methods.

\begin{figure}[htp]
\centering
\subfloat{\includegraphics[trim=1.7cm 1.3cm 3cm 0.9cm, clip=true, width=1.15in]{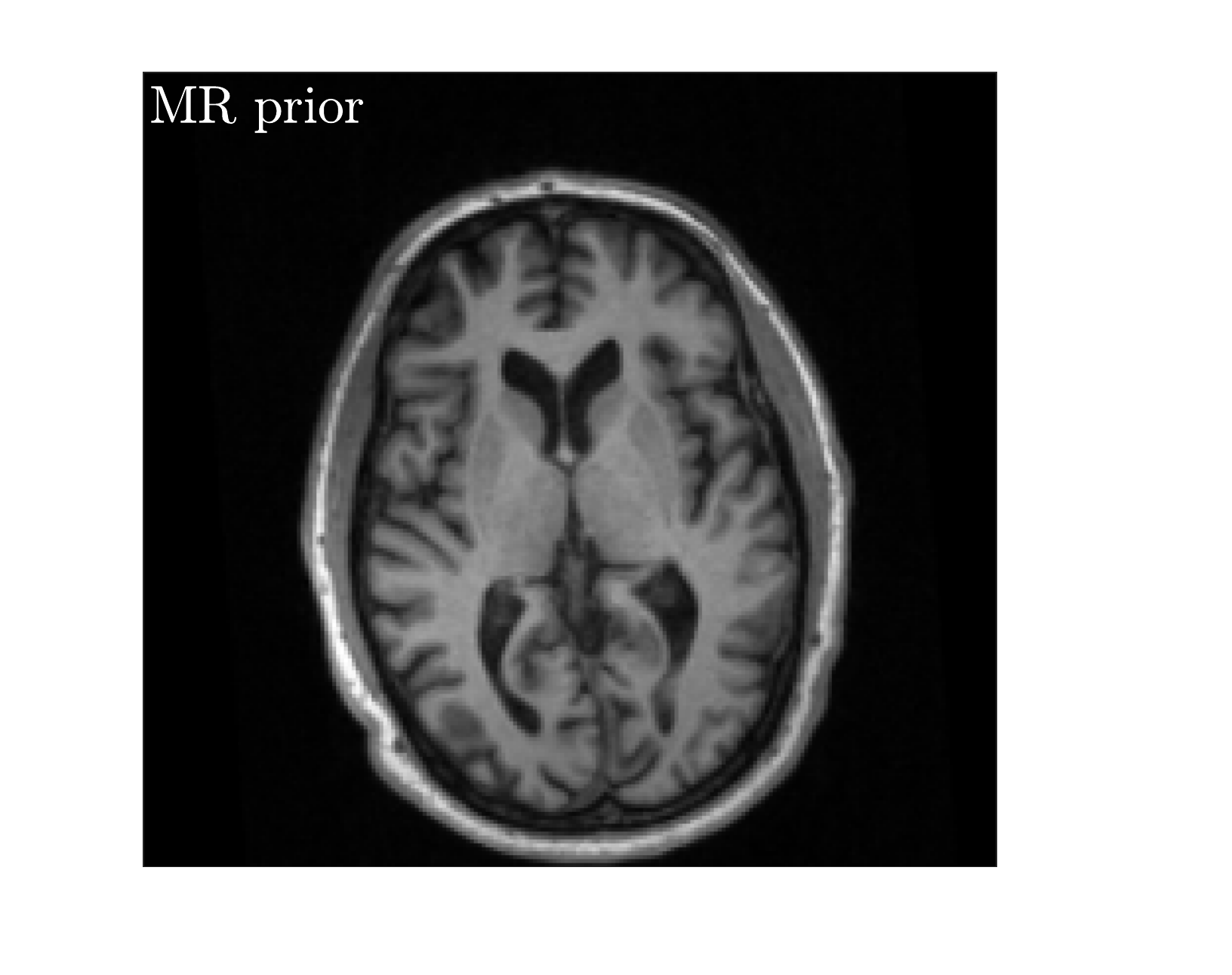}}  
\subfloat{\includegraphics[trim=5.6cm 4.5cm 9.5cm 3.8cm, clip=true, width=4.5in]{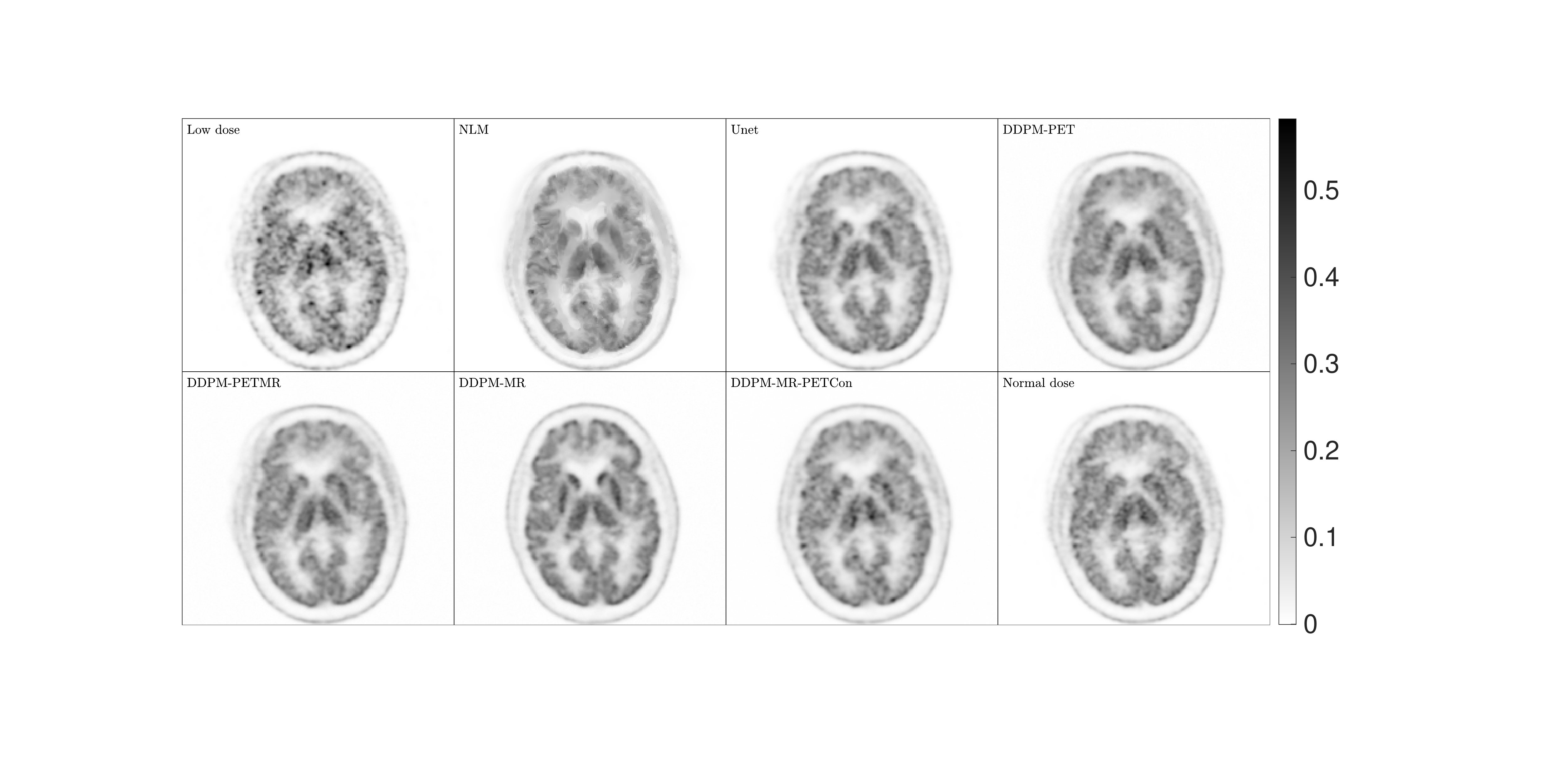}} \\
\subfloat{\includegraphics[trim=1.7cm 1.cm 3cm 0.9cm, clip=true, width=1.15in]{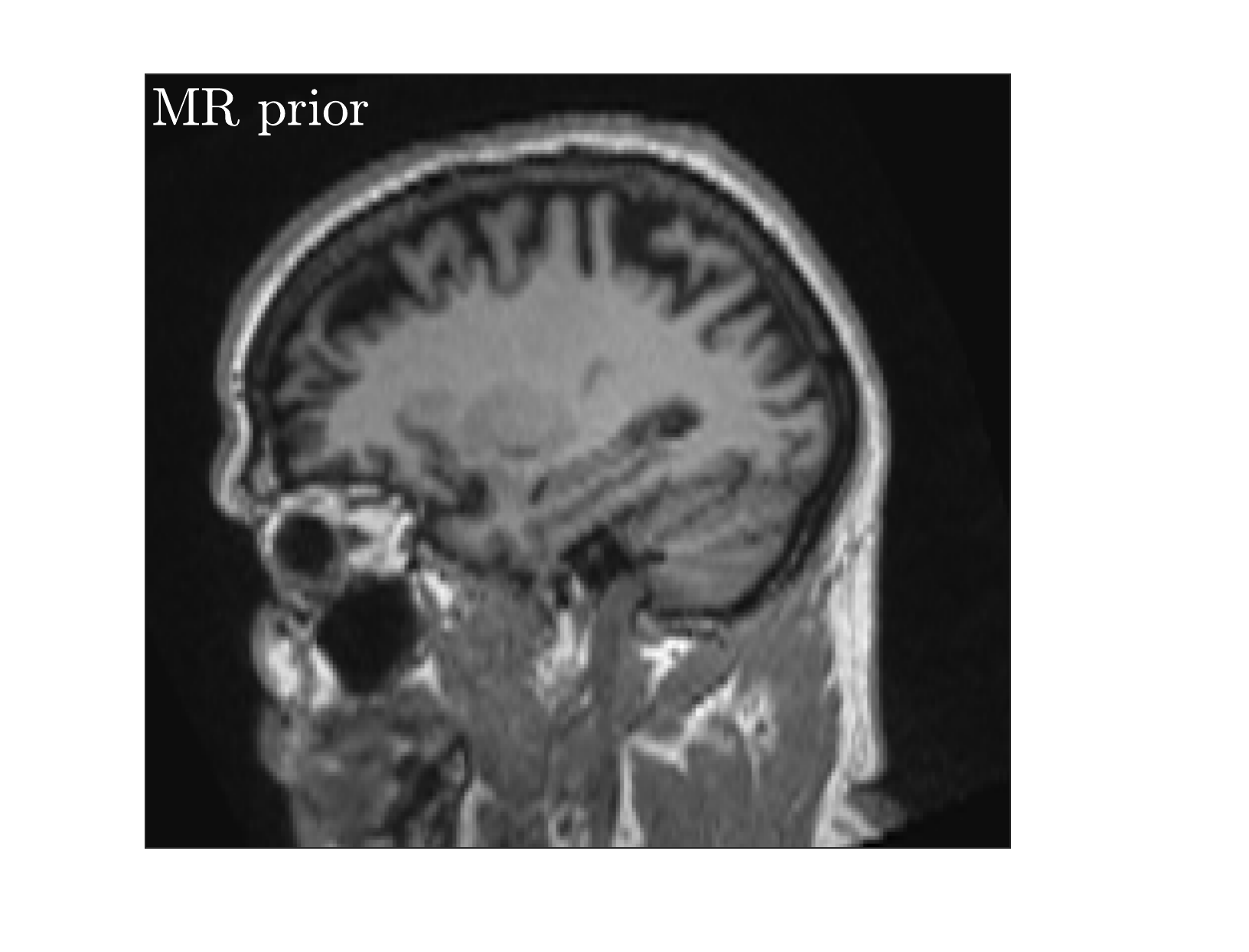}}  
\subfloat{\includegraphics[trim=5.6cm 4.5cm 9.5cm 4.1cm, clip=true, width=4.5in]{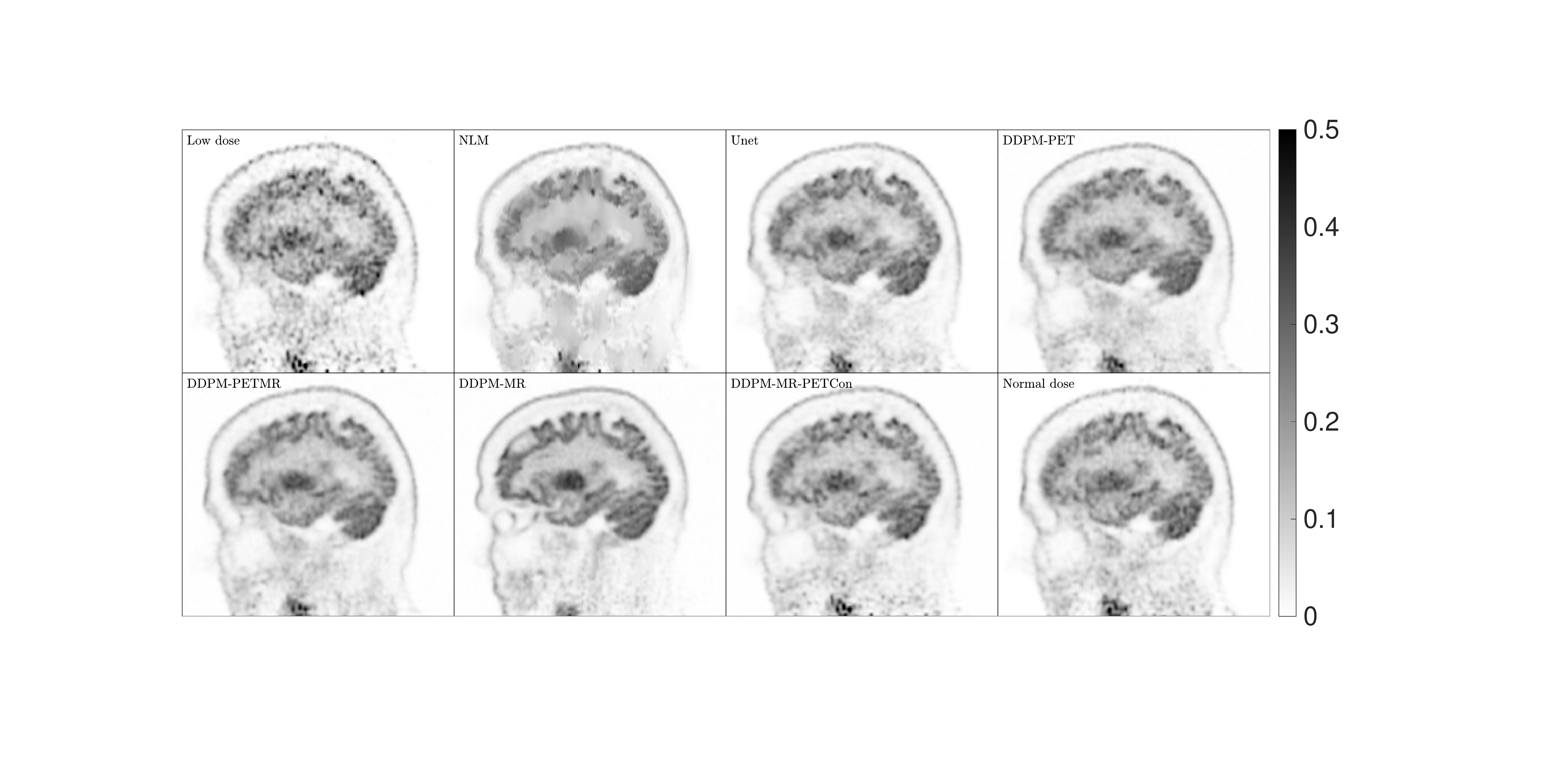}}  \\
\subfloat{\includegraphics[trim=1.7cm 1.cm 3cm 0.9cm, clip=true, width=1.15in]{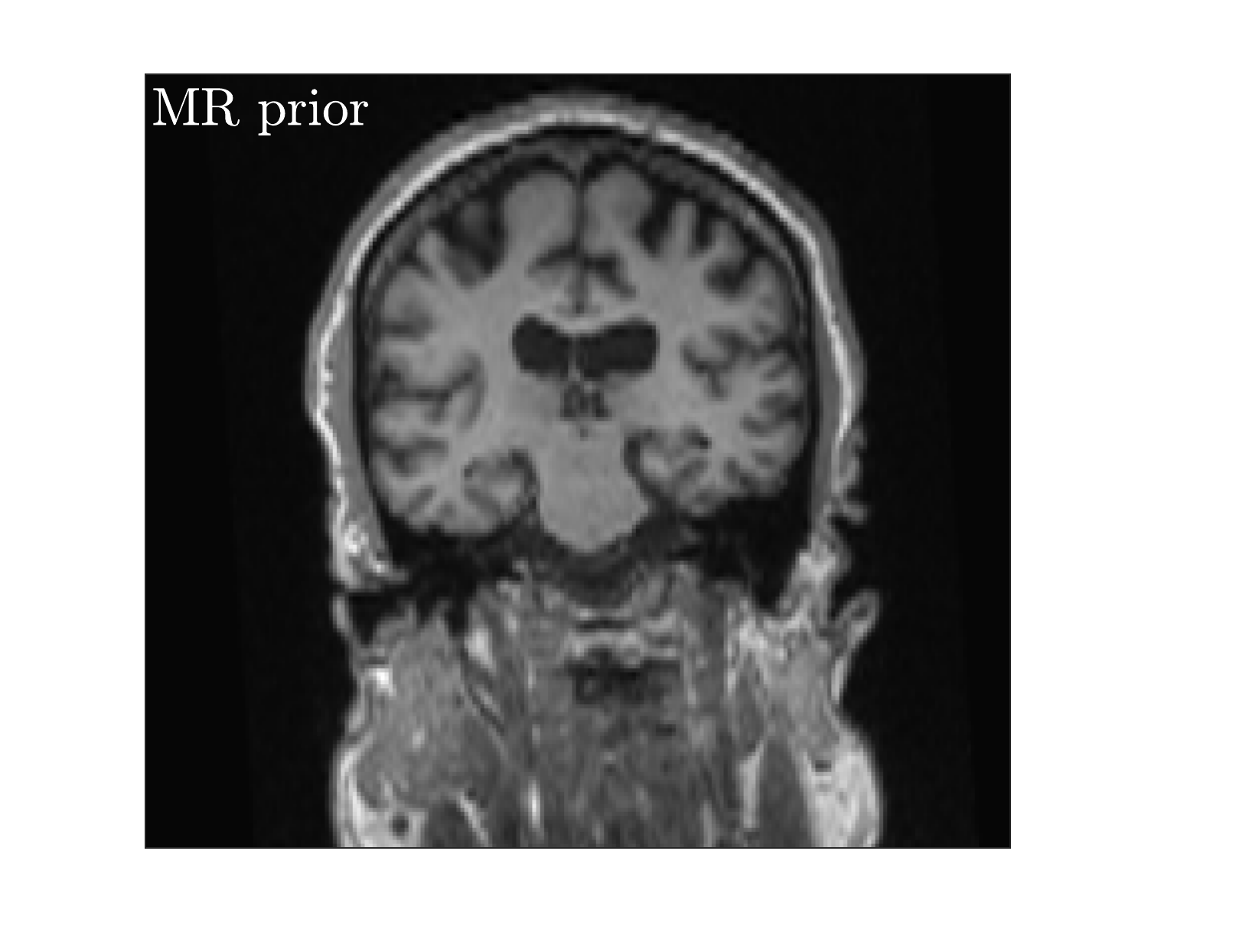}}  
\subfloat{\includegraphics[trim=5.6cm 4.5cm 9.5cm 4.1cm, clip=true, width=4.5in]{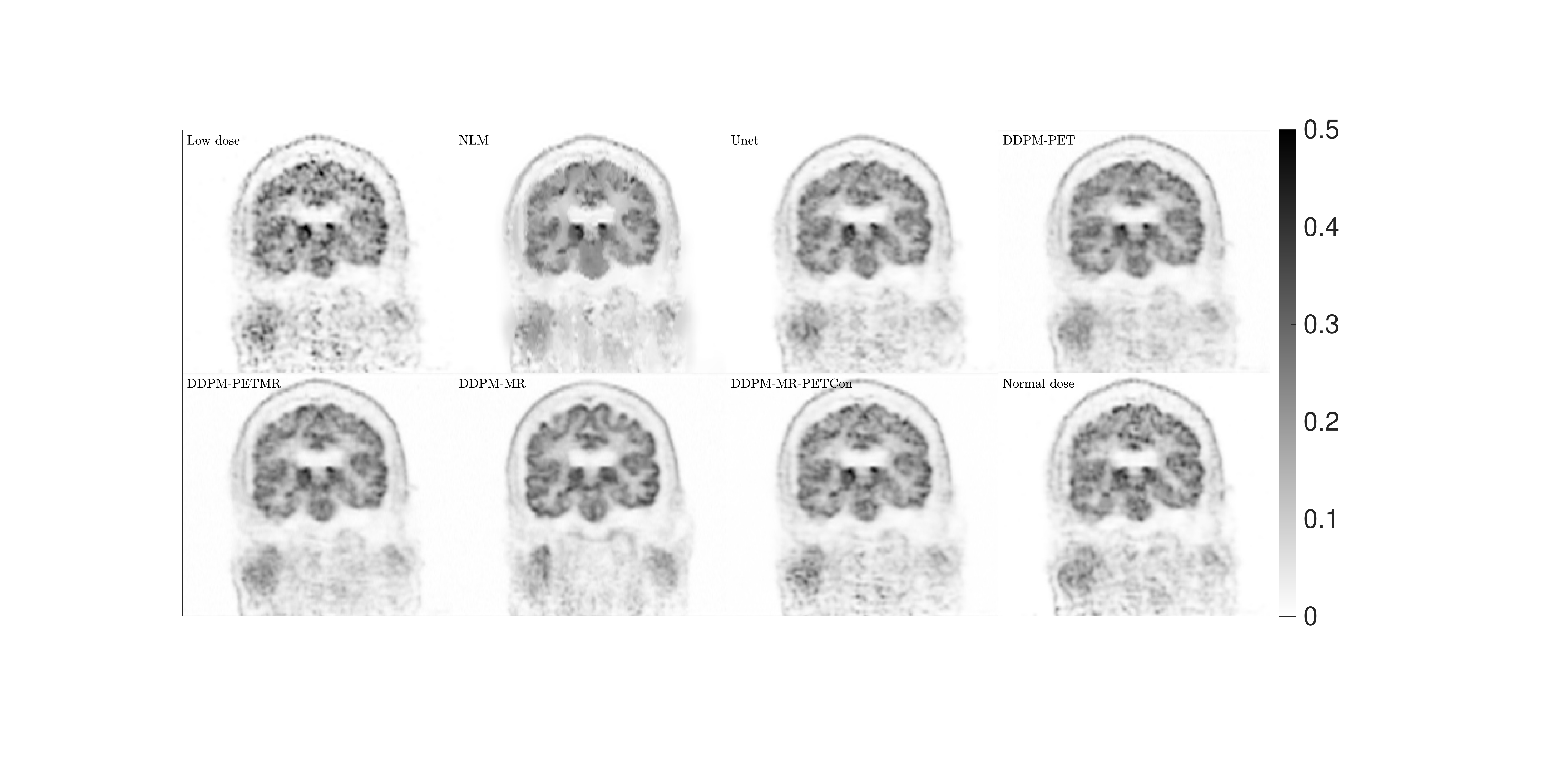}} 
\caption{Three views of one $^{18}$F-MK-6240 test dataset processed using different methods along with the MR prior image (1st column),  the 1/4 low-dose image (2nd column) and the normal-dose image (last column).  }
\label{fig:mk_img}
\end{figure}

\noindent Fig.~\ref{fig:mk_img} shows three views of one  $^{18}$F-MK-6240 dataset processed using different methods along with the co-registered MR prior image.  Compared to DDPM-PET,  DDPM-PETMR has higher image resolution,  which demonstrates the benefit of including the additional MR prior image as the network input.   It can also be observed that the DDPM-MR method can generate results with the highest resolution and the lowest noise.  However,  large bias exists at some regions,  e.g., the caudate and putamen regions.  Based on the additional PET data-consistency constraint,  results of the DDPM-MR-PETCon method share more similarity with the normal-dose results.  Compared to Unet,  DDPM-MR-PETCon has higher image contrast.  Fig.~\ref{fig:mk_psnr} shows the PSNR and SSIM values of different methods based on 20 test datasets.  These global quantification values show that DDPM-PETMR has the best performance, followed by DDPM-MR-PETCon.  The regional PSNR plot at 14 cortical regions for different methods is presented at Fig.~\ref{fig:mk_local_psnr}.  Fig.~\ref{fig:mk_surface} further shows the surface plots of the left hemisphere regarding the PET relative error for different methods.  The regional and the surface results both show that DDPM-MR-PETCon has the lowest error, followed by the DDPM-PETMR method.  It should be noted that the DDPM-MR-PETCon method does not have specific noise-level requirement, and it can be applied to different noise levels,  which is more flexible than DDPM-PETMR.

\begin{figure}[htp]
\centering
\subfloat{\includegraphics[trim=5.9cm 8.4cm 5.9cm 7.8cm, clip=true, width=6in]{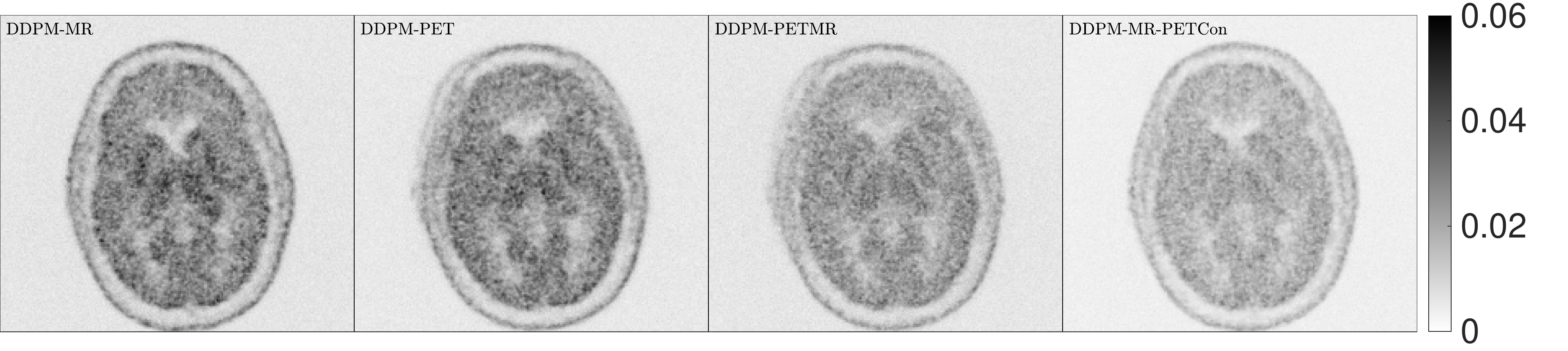}}  \\[-0.07cm]
\subfloat{\includegraphics[trim=5.9cm 8.4cm 5.9cm 7.8cm, clip=true, width=6in]{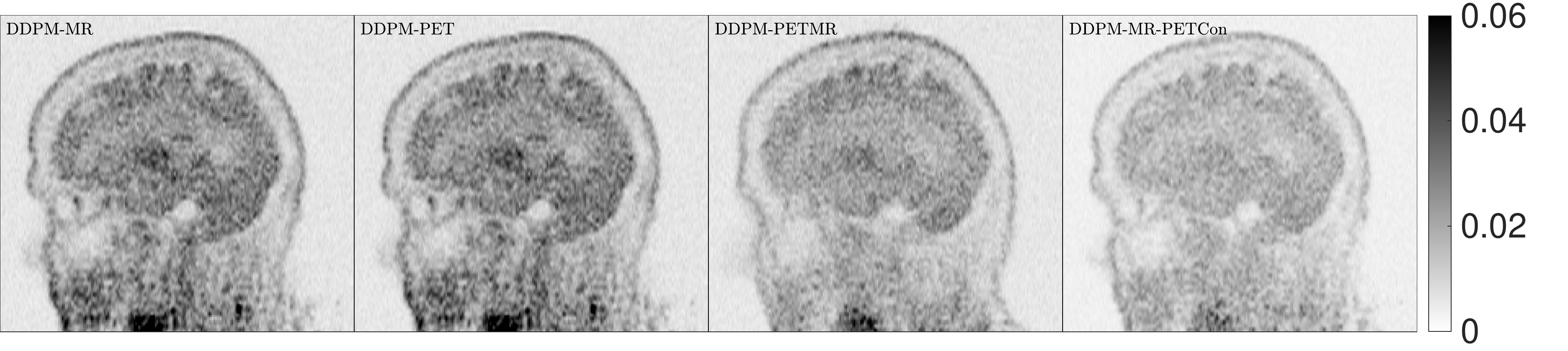}}  \\[-0.07cm]
\subfloat{\includegraphics[trim=5.9cm 8.4cm 5.9cm 7.8cm, clip=true, width=6in]{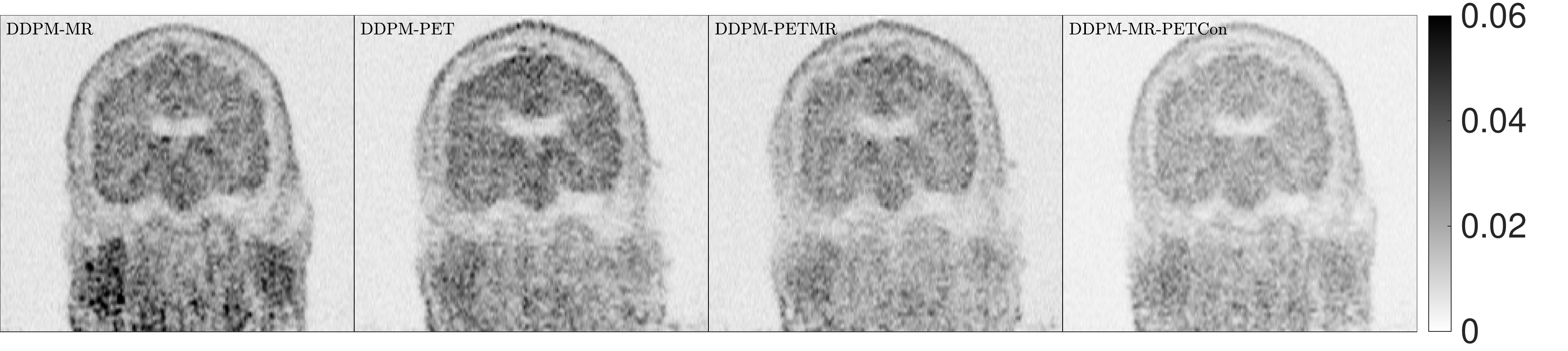}} 
\caption{Three views of the uncertainty maps of one $^{18}$F-MK-6240 test dataset calculated from 20 realizations for different DDPM-based methods.}
\label{fig:mk_variance}
\end{figure}

\noindent One characteristic of the diffusion models is the ability to generate the uncertainty map based on multiple realizations.   Fig.~\ref{fig:mk_variance} shows the uncertainty maps of the DDPM-MR, DDPM-PET, DDPM-PETMR and DDPM-MR-PETCon methods based on 20 realizations from the same dataset as in Fig.~\ref{fig:mk_img}.  Comparing the results of DDPM-PET and DDPM-PETMR,  we can observe that adding additional MR prior image can reduce the uncertainty.  Comparing the results of DDPM-MR and DDPM-MR-PETCon,  we can see that adding PET data-consistency constraint can further reduce the uncertainty.  Finally,  the uncertainty values of DDPM-MR-PETCon are smaller than that of DDPM-PETMR,  which means including the PET image during the refinement steps can further reduce the uncertainty compared to supplying it as the network input.

\section*{Discussion}

In this work we developed DDPM-based approaches for PET image denoising and evaluated their performance  on datasets from two different tracers.  Quantitative results show that the DDPM-based frameworks with PET information included can have better performance than the NLM and Unet-based denoising methods.  One benefit of the DDPM-based framework compared to Unet is the ability to calculate the uncertainty map by generating multiple realizations,  which has the potential to be used in progression tracking and longitudinal studies.

\begin{figure}[htp]
\centering
\subfloat{\includegraphics[trim=0cm 0.8cm 0cm 0cm, clip=true, width=7in]{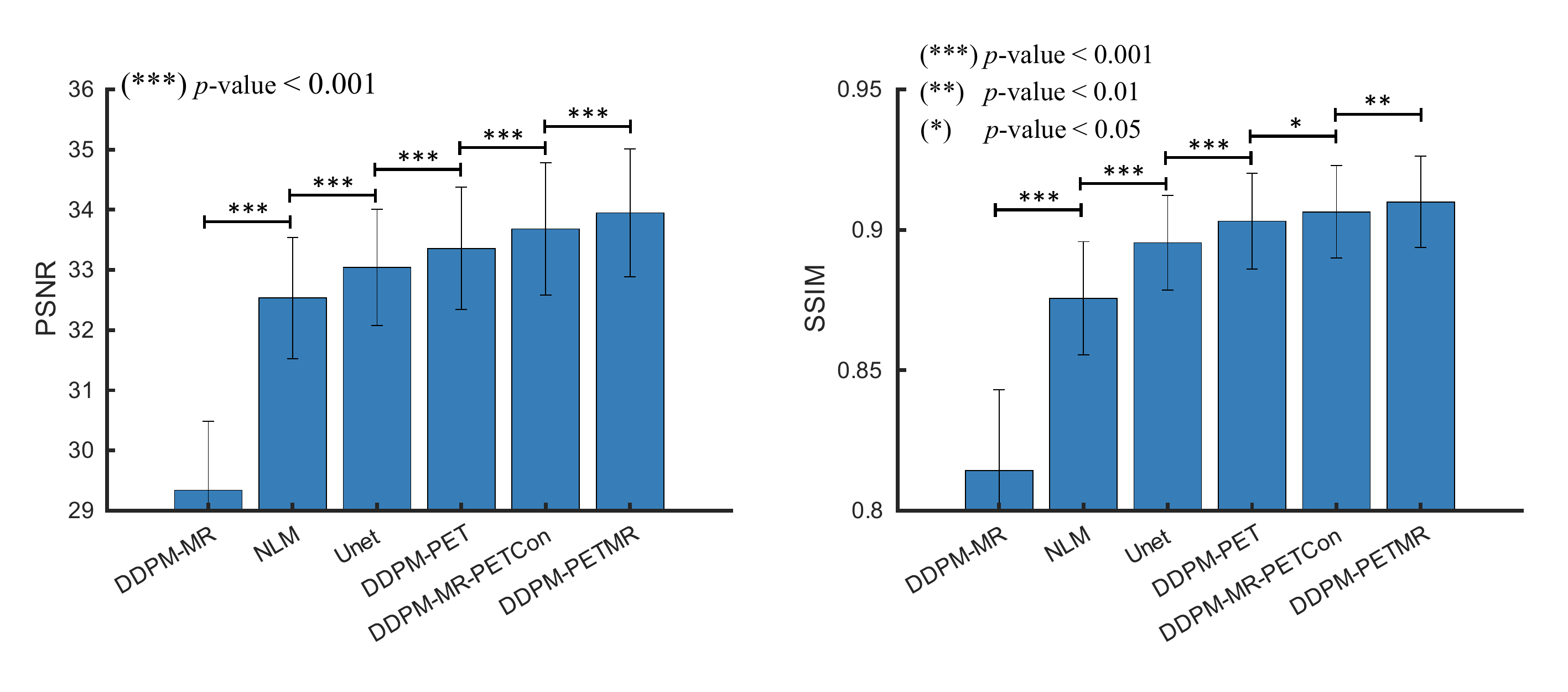}} 
\caption{The PSNR (left) and SSIM (right) values using different methods calculated based on 20 $^{18}$F-MK-6240 test datasets.  *, **, *** located at the top of the bar plot represent $p$-value$<0.05$, $p$-value$<0.01$, and $p$-value$<0.001$, respectively.}
\label{fig:mk_psnr}
\end{figure}

\begin{figure*}[htp]
\centering
\subfloat{\includegraphics[trim=4cm 0.2cm 1.2cm 0.5cm, clip=true, width=7.5in]{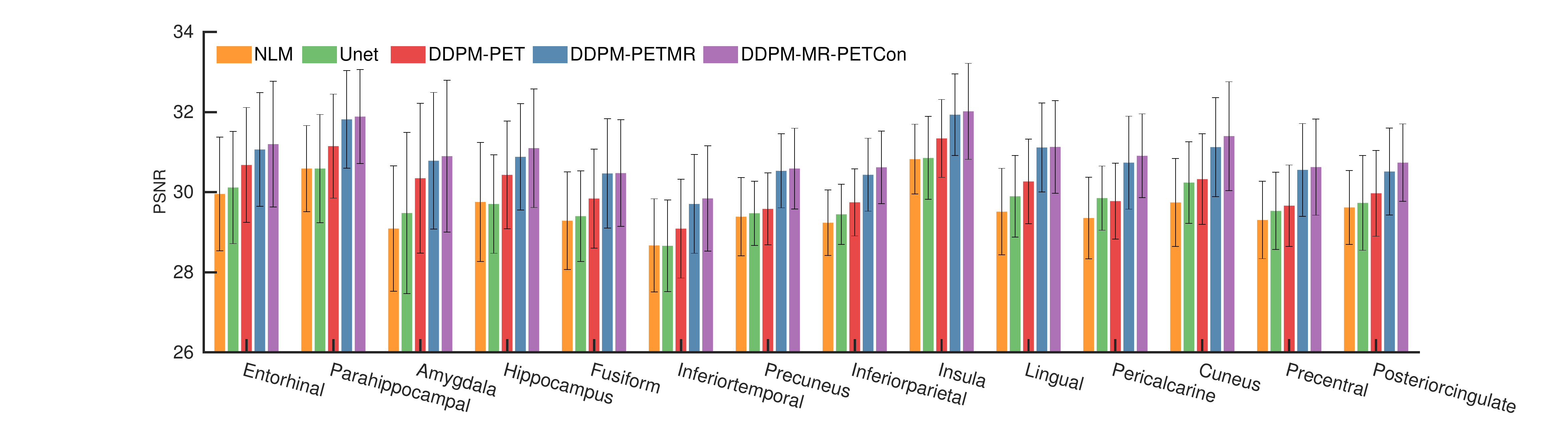}}  \\[-0.07cm]
\caption{ PSNR values of the 14 cortical regions for different methods calculated based on 20 $^{18}$F-MK-6240 test datasets. }
\label{fig:mk_local_psnr}
\end{figure*}

\begin{figure*}[htp]
\centering
\subfloat{\includegraphics[trim=0cm 0.cm 0cm 0cm, clip=true, width=7.5in]{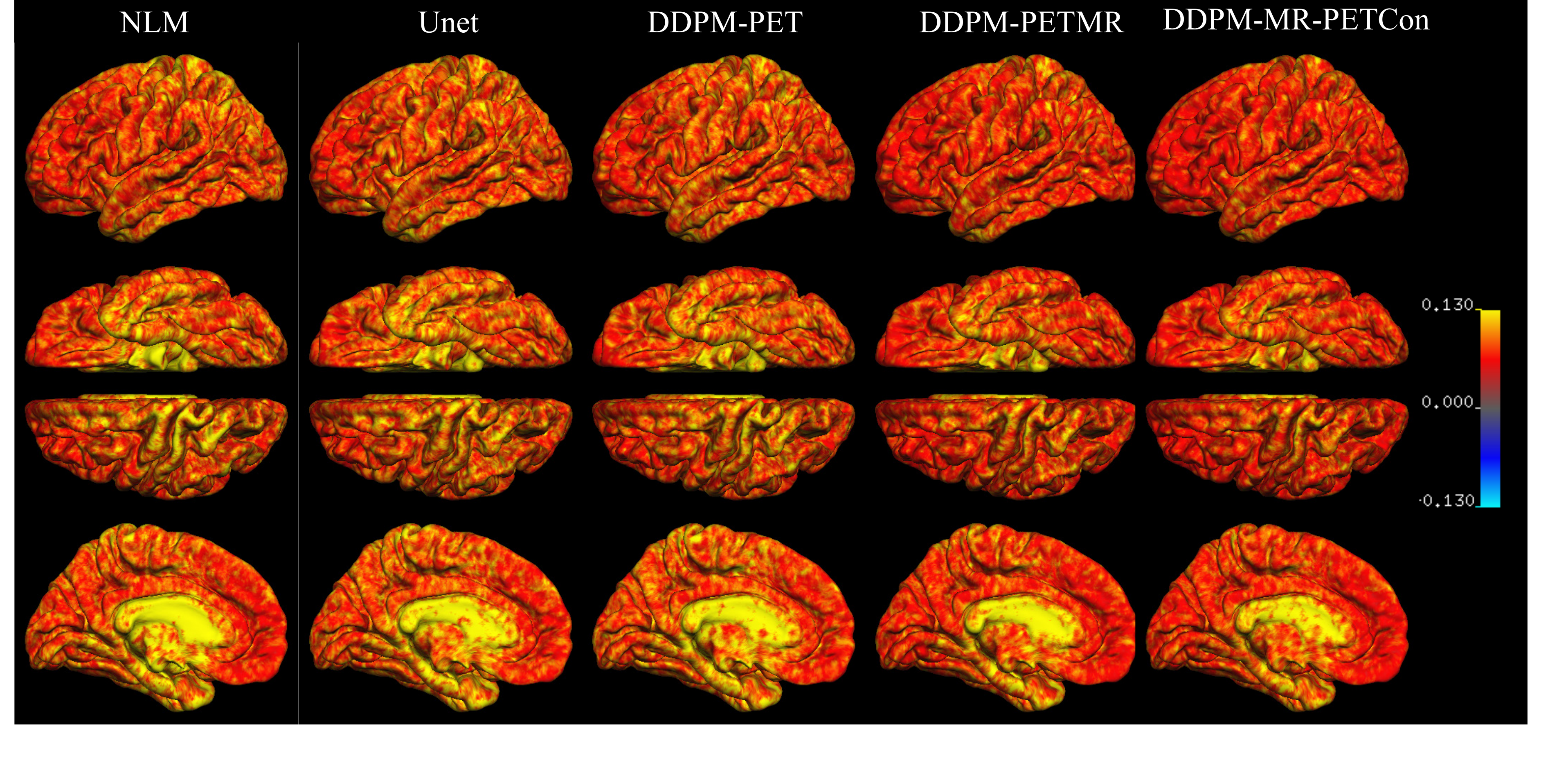}} 
\caption{Surface maps of the left hemisphere regarding mean PET relative errors for different methods calculated based on 20 $^{18}$F-MK-6240 test datasets. }
\label{fig:mk_surface}
\end{figure*}

\noindent Regarding different DDPM-based methods,  by comparing DDPM-PET and DDPM-PETMR, our evaluations show that further adding the MR prior images as the additional network input can help improve the model's performance both globally (as shown in Fig.~\ref{fig:mk_psnr}) and regionally (as shown in Figs.~\ref{fig:mk_local_psnr} and~\ref{fig:mk_surface}).  It can also further reduce the uncertainty values as shown in Fig.~\ref{fig:mk_variance}.  The results of DDPM-MR presented in Figs.~\ref{fig:mk_img} and ~\ref{fig:mk_psnr} show that only replying on the MR prior image is difficult to synthesize a pseudo-PET image with high accuracy as large bias can exist in the results due to the high dynamic range of PET uptakes in different regions.  Further adding PET information as a data-consistency constraint in the inference step can help improve the results as shown by the performance of DDPM-MR-PETCon.  

\noindent Fig.~\ref{fig:mk_psnr} shows that DDPM-PETMR has better performance than DDPM-MR-PETCon regarding global quantification.  Figs.~\ref{fig:mk_local_psnr} and~\ref{fig:mk_surface} reveal that DDPM-MR-PETCon is better than DDPM-PETMR with respect to local quantification.  One advantage of DDPM-MR-PETCon compared to DDPM-PETMR and Unet is that it can be applied to PET images of different noise levels.  This is because during the training phase of the score function $\bb{\epsilon}_{\bb{\theta}}$,  PET information is not included.  Another advantage of DDPM-MR-PETCon is that it can be further extended to PET image reconstruction. This can be achieved by changing the Gaussian distribution assumption in equation~(\ref{eq:data_constraint}) to Poisson distribution with PET sinograms included \cite{qi2006iterative}.  Considering the performance,  the invariability to nosie levels, and the potential extension to PET image reconstruction,  the DDPM-MR-PETCon framework is preferable over DDPM-PETMR.

\noindent In this work,   a 2D network was used in the DDPM-based methods and the neighboring two axial slices were supplied as the network input to reduce axial artifacts.  We have tested the 3D network and the training is 4 times slower than the 2D network, which is not realistic based on our current computing resources.  Additionally,  the inference step took around 50 mins for one 3D dataset.  This is one issue with the diffusion models and a lot of on-going works are trying to reduce the inference time, e.g., finding a better initialization instead of starting from the random noise \cite{chung2022come}.   Further extending the framework to enable efficient 3D network training and reduce the inference time is one direction of our future work.  The current study is based on brain datasets,  and further evaluating the performance of the DDPM-based methods on whole-body PET datasets of different tracers is another direction of our future work.

\section*{Conclusion}
In this work,  DDPM-based PET imaging denoising frameworks were proposed and evaluated based on $^{18}$F-FDG and  $^{18}$F-MK-6240 datasets.  Results show that DDPM-based frameworks including PET information have better performance than the nonlocal mean and the Unet-based denoising methods.  Adding additional MR prior in the model can  achieve better performance and further reduce the uncertainty during image denoising.  Future work will focus on extending the framework to 3D network,  further reducing the inference time,  and performing more evaluations using whole-body PET datasets of different tracers.


\clearpage
\bibliographystyle{ieeetr}

\clearpage

\section*{Declarations}

\noindent \textbf{Funding} This work was supported by the National Institutes of Health under grants R21AG067422,  R03EB030280, R01AG078250, P41EB022544 and P01AG036694. 

\noindent \textbf{Competing Interests} The authors have no relevant financial or non-financial interests to disclose.. 

\noindent \textbf{Ethics approval} All procedures performed in studies involving human participants were in accordance with the ethical standards of the institutional and/or national research committee and with the 1964 Helsinki declaration and its later amendments or comparable ethical standards. 

\noindent \textbf{Informed consent } For the ${^{18}}$F-FDG datasets, informed consent was waived due to the retrospective merits of the datasets.  For the ${^{18}}$F-MK-6240 datasets, informed consent was obtained from all the participants.

\clearpage
\newpage

\clearpage
\newpage

\clearpage
\newpage

\clearpage
\newpage


\clearpage
\newpage

\clearpage
\newpage

\clearpage
\newpage

\clearpage
\newpage

\end{document}

%% file: defs.tex
\makeatletter
\renewcommand{\@biblabel}[1]{[#1]}
\def\baselinestretch{2}
\renewcommand\section{\@startsection {section}{1}{\z@}%
                                   {-3.5ex \@plus -1ex \@minus -.2ex}%
                                   {2.3ex \@plus.2ex}%
                                   {\normalfont\large\bfseries}}

\renewcommand\subsection{\@startsection{subsubsection}{3}{0pt}%
                                   {-\medskipamount}%
                                   {8pt}%
				   {\normalfont\it}}

\renewcommand\subsubsection{\@startsection{subsubsection}{3}{0pt}%
                                   {-\medskipamount}%
                                   {8pt}%
                                   {\normalfont}}
\makeatother

\makeatletter
\renewcommand\@biblabel[1]{#1.}     
\def\@cite#1#2{({#1\if@tempswa , #2\fi})}
\def\@eqnnum{{\normalfont \normalcolor [\theequation]}} 
\def\eqref#1{[\ref{#1}]}
\makeatother

\makeatletter
\def\maketitle{%
  \null
  \thispagestyle{empty}%
  \vfill
  \begin{center}\leavevmode
    \normalfont
    {\Large \@title\par}%
    \vskip 1cm
    {\large \@author\par}%
  \end{center}%
  \vskip 1cm
\renewcommand{\baselinestretch}{1.2}
{\normalsize \begin{tabular*}{\linewidth}{lp{7in}}
    \quad  $^1$ Gordon Center for Medical Imaging,  \\
     \quad Massachusetts General Hospital, Harvard Medical School, Massachusetts, USA \\ 
     \quad  $^2$ Center for Advanced Medical Computing and Analysis,  \\
     \quad Massachusetts General Hospital, Harvard Medical School, Massachusetts, USA \\      
     \quad  $^3$ Department of Imaging Physics, \\
     \quad  University of Texas MD Anderson Cancer Center, Texas, USA  \\      
  \end{tabular*}
}

  \vskip 0.5cm
  Correspondence to: \vspace{0.2cm} \\
  \renewcommand{\baselinestretch}{1.2}
  {\normalsize
  \indent Kuang Gong, Ph.D.

  \indent Massachusetts General Hospital

  \indent 125 Nashua Street Suite 660, Boston, MA, 02114, USA

  \indent E-mail: kgong@mgh.harvard.edu
}
  \vskip 1.0cm
  \vfill
  \null
\renewcommand{\baselinestretch}{1.5}
  \cleardoublepage
  }
\makeatother

%% file: DDPM_newformat_v2.3.bbl
\begin{thebibliography}{10}

\bibitem{lin2001improving}
J.-W. Lin, A.~F. Laine, and S.~R. Bergmann, ``Improving pet-based physiological
  quantification through methods of wavelet denoising,'' {\em IEEE Transactions
  on Biomedical Engineering}, vol.~48, no.~2, pp.~202--212, 2001.

\bibitem{christian2010dynamic}
B.~T. Christian, N.~T. Vandehey, J.~M. Floberg, and C.~A. Mistretta, ``Dynamic
  pet denoising with hypr processing,'' {\em Journal of Nuclear Medicine},
  vol.~51, no.~7, pp.~1147--1154, 2010.

\bibitem{chan2014postreconstruction}
C.~Chan, R.~Fulton, R.~Barnett, D.~D. Feng, and S.~Meikle, ``Postreconstruction
  nonlocal means filtering of whole-body pet with an anatomical prior,'' {\em
  IEEE Transactions on medical imaging}, vol.~33, no.~3, pp.~636--650, 2014.

\bibitem{dutta2013non}
J.~Dutta, R.~M. Leahy, and Q.~Li, ``Non-local means denoising of dynamic pet
  images,'' {\em PloS one}, vol.~8, no.~12, p.~e81390, 2013.

\bibitem{yan2015mri}
J.~Yan, J.~C.-S. Lim, and D.~W. Townsend, ``Mri-guided brain pet image
  filtering and partial volume correction,'' {\em Physics in medicine and
  biology}, vol.~60, no.~3, p.~961, 2015.

\bibitem{ote2020kinetics}
K.~Ote, F.~Hashimoto, A.~Kakimoto, T.~Isobe, T.~Inubushi, R.~Ota, A.~Tokui,
  A.~Saito, T.~Moriya, T.~Omura, {\em et~al.}, ``Kinetics-induced block
  matching and 5-d transform domain filtering for dynamic pet image
  denoising,'' {\em IEEE Transactions on Radiation and Plasma Medical
  Sciences}, vol.~4, no.~6, pp.~720--728, 2020.

\bibitem{wang2016semisupervised}
Y.~Wang, G.~Ma, L.~An, F.~Shi, P.~Zhang, D.~S. Lalush, X.~Wu, Y.~Pu, J.~Zhou,
  and D.~Shen, ``Semisupervised tripled dictionary learning for standard-dose
  pet image prediction using low-dose pet and multimodal mri,'' {\em IEEE
  transactions on biomedical engineering}, vol.~64, no.~3, pp.~569--579, 2016.

\bibitem{xiang2017deep}
L.~Xiang, Y.~Qiao, D.~Nie, L.~An, W.~Lin, Q.~Wang, and D.~Shen, ``Deep
  auto-context convolutional neural networks for standard-dose pet image
  estimation from low-dose pet/mri,'' {\em Neurocomputing}, vol.~267,
  pp.~406--416, 2017.

\bibitem{kaplan2019full}
S.~Kaplan and Y.-M. Zhu, ``Full-dose pet image estimation from low-dose pet
  image using deep learning: a pilot study,'' {\em Journal of digital imaging},
  vol.~32, no.~5, pp.~773--778, 2019.

\bibitem{cui2019pet}
J.~Cui, K.~Gong, N.~Guo, C.~Wu, X.~Meng, K.~Kim, K.~Zheng, Z.~Wu, L.~Fu, B.~Xu,
  {\em et~al.}, ``Pet image denoising using unsupervised deep learning,'' {\em
  European journal of nuclear medicine and molecular imaging}, vol.~46, no.~13,
  pp.~2780--2789, 2019.

\bibitem{chen2019ultra}
K.~T. Chen, E.~Gong, F.~B. de~Carvalho~Macruz, J.~Xu, A.~Boumis, M.~Khalighi,
  K.~L. Poston, S.~J. Sha, M.~D. Greicius, E.~Mormino, {\em et~al.},
  ``Ultra--low-dose 18f-florbetaben amyloid pet imaging using deep learning
  with multi-contrast mri inputs,'' {\em Radiology}, vol.~290, no.~3,
  pp.~649--656, 2019.

\bibitem{hashimoto2019dynamic}
F.~Hashimoto, H.~Ohba, K.~Ote, A.~Teramoto, and H.~Tsukada, ``Dynamic pet image
  denoising using deep convolutional neural networks without prior training
  datasets,'' {\em IEEE access}, vol.~7, pp.~96594--96603, 2019.

\bibitem{da2020micro}
C.~O. da~Costa-Luis and A.~J. Reader, ``Micro-networks for robust mr-guided low
  count pet imaging,'' {\em IEEE transactions on radiation and plasma medical
  sciences}, vol.~5, no.~2, pp.~202--212, 2020.

\bibitem{schramm2021approximating}
G.~Schramm, D.~Rigie, T.~Vahle, A.~Rezaei, K.~Van~Laere, T.~Shepherd, J.~Nuyts,
  and F.~Boada, ``Approximating anatomically-guided pet reconstruction in image
  space using a convolutional neural network,'' {\em Neuroimage}, vol.~224,
  p.~117399, 2021.

\bibitem{mehranian2022image}
A.~Mehranian, S.~D. Wollenweber, M.~D. Walker, K.~M. Bradley, P.~A. Fielding,
  K.-H. Su, R.~Johnsen, F.~Kotasidis, F.~P. Jansen, and D.~R. McGowan, ``Image
  enhancement of whole-body oncology [18f]-fdg pet scans using deep neural
  networks to reduce noise,'' {\em European journal of nuclear medicine and
  molecular imaging}, vol.~49, no.~2, pp.~539--549, 2022.

\bibitem{daveau2022deep}
R.~S. Daveau, I.~Law, O.~M. Henriksen, S.~G. Hasselbalch, U.~B. Andersen,
  L.~Anderberg, L.~H{\o}jgaard, F.~L. Andersen, and C.~N. Ladefoged, ``Deep
  learning based low-activity pet reconstruction of [11c] pib and [18f] fe-pe2i
  in neurodegenerative disorders,'' {\em NeuroImage}, vol.~259, p.~119412,
  2022.

\bibitem{ouyang2019ultra}
J.~Ouyang, K.~T. Chen, E.~Gong, J.~Pauly, and G.~Zaharchuk, ``Ultra-low-dose
  pet reconstruction using generative adversarial network with feature matching
  and task-specific perceptual loss,'' {\em Medical physics}, vol.~46, no.~8,
  pp.~3555--3564, 2019.

\bibitem{lei2019whole}
Y.~Lei, X.~Dong, T.~Wang, K.~Higgins, T.~Liu, W.~J. Curran, H.~Mao, J.~A. Nye,
  and X.~Yang, ``Whole-body pet estimation from low count statistics using
  cycle-consistent generative adversarial networks,'' {\em Physics in Medicine
  \& Biology}, vol.~64, no.~21, p.~215017, 2019.

\bibitem{zhou2020supervised}
L.~Zhou, J.~D. Schaefferkoetter, I.~W. Tham, G.~Huang, and J.~Yan, ``Supervised
  learning with cyclegan for low-dose fdg pet image denoising,'' {\em Medical
  image analysis}, vol.~65, p.~101770, 2020.

\bibitem{song2020pet}
T.-A. Song, S.~R. Chowdhury, F.~Yang, and J.~Dutta, ``Pet image
  super-resolution using generative adversarial networks,'' {\em Neural
  Networks}, vol.~125, pp.~83--91, 2020.

\bibitem{sanaat2021deep}
A.~Sanaat, I.~Shiri, H.~Arabi, I.~Mainta, R.~Nkoulou, and H.~Zaidi, ``Deep
  learning-assisted ultra-fast/low-dose whole-body pet/ct imaging,'' {\em
  European journal of nuclear medicine and molecular imaging}, vol.~48, no.~8,
  pp.~2405--2415, 2021.

\bibitem{xue2022cross}
S.~Xue, R.~Guo, K.~P. Bohn, J.~Matzke, M.~Viscione, I.~Alberts, H.~Meng,
  C.~Sun, M.~Zhang, M.~Zhang, {\em et~al.}, ``A cross-scanner and cross-tracer
  deep learning method for the recovery of standard-dose imaging quality from
  low-dose pet,'' {\em European journal of nuclear medicine and molecular
  imaging}, vol.~49, no.~6, pp.~1843--1856, 2022.

\bibitem{gong2018pet}
K.~Gong, J.~Guan, C.-C. Liu, and J.~Qi, ``Pet image denoising using a deep
  neural network through fine tuning,'' {\em IEEE Transactions on Radiation and
  Plasma Medical Sciences}, vol.~3, no.~2, pp.~153--161, 2018.

\bibitem{liu2020noise}
H.~Liu, J.~Wu, W.~Lu, J.~A. Onofrey, Y.-H. Liu, and C.~Liu, ``Noise reduction
  with cross-tracer and cross-protocol deep transfer learning for low-dose
  pet,'' {\em Physics in Medicine \& Biology}, vol.~65, no.~18, p.~185006,
  2020.

\bibitem{chen2020generalization}
K.~T. Chen, M.~Sch{\"u}rer, J.~Ouyang, M.~E.~I. Koran, G.~Davidzon, E.~Mormino,
  S.~Tiepolt, K.-T. Hoffmann, O.~Sabri, G.~Zaharchuk, {\em et~al.},
  ``Generalization of deep learning models for ultra-low-count amyloid pet/mri
  using transfer learning,'' {\em European journal of nuclear medicine and
  molecular imaging}, vol.~47, no.~13, pp.~2998--3007, 2020.

\bibitem{cui2021populational}
J.~Cui, K.~Gong, N.~Guo, C.~Wu, K.~Kim, H.~Liu, and Q.~Li, ``Populational and
  individual information based pet image denoising using conditional
  unsupervised learning,'' {\em Physics in Medicine \& Biology}, vol.~66,
  no.~15, p.~155001, 2021.

\bibitem{zhou2022federated}
B.~Zhou, T.~Miao, N.~Mirian, X.~Chen, H.~Xie, Z.~Feng, X.~Guo, X.~Li, S.~K.
  Zhou, J.~S. Duncan, {\em et~al.}, ``Federated transfer learning for low-dose
  pet denoising: A pilot study with simulated heterogeneous data,'' {\em IEEE
  Transactions on Radiation and Plasma Medical Sciences}, 2022.

\bibitem{ho2020denoising}
J.~Ho, A.~Jain, and P.~Abbeel, ``Denoising diffusion probabilistic models,''
  {\em Advances in Neural Information Processing Systems}, vol.~33,
  pp.~6840--6851, 2020.

\bibitem{song2019generative}
Y.~Song and S.~Ermon, ``Generative modeling by estimating gradients of the data
  distribution,'' {\em Advances in Neural Information Processing Systems},
  vol.~32, 2019.

\bibitem{song2020score}
Y.~Song, J.~Sohl-Dickstein, D.~P. Kingma, A.~Kumar, S.~Ermon, and B.~Poole,
  ``Score-based generative modeling through stochastic differential
  equations,'' {\em arXiv preprint arXiv:2011.13456}, 2020.

\bibitem{dhariwal2021diffusion}
P.~Dhariwal and A.~Nichol, ``Diffusion models beat gans on image synthesis,''
  {\em Advances in Neural Information Processing Systems}, vol.~34,
  pp.~8780--8794, 2021.

\bibitem{rombach2022high}
R.~Rombach, A.~Blattmann, D.~Lorenz, P.~Esser, and B.~Ommer, ``High-resolution
  image synthesis with latent diffusion models,'' in {\em Proceedings of the
  IEEE/CVF Conference on Computer Vision and Pattern Recognition},
  pp.~10684--10695, 2022.

\bibitem{saharia2021image}
C.~Saharia, J.~Ho, W.~Chan, T.~Salimans, D.~J. Fleet, and M.~Norouzi, ``Image
  super-resolution via iterative refinement,'' {\em arXiv preprint
  arXiv:2104.07636}, 2021.

\bibitem{lugmayr2022repaint}
A.~Lugmayr, M.~Danelljan, A.~Romero, F.~Yu, R.~Timofte, and L.~Van~Gool,
  ``Repaint: Inpainting using denoising diffusion probabilistic models,'' in
  {\em Proceedings of the IEEE/CVF Conference on Computer Vision and Pattern
  Recognition}, pp.~11461--11471, 2022.

\bibitem{jalal2021robust}
A.~Jalal, M.~Arvinte, G.~Daras, E.~Price, A.~G. Dimakis, and J.~Tamir, ``Robust
  compressed sensing mri with deep generative priors,'' {\em Advances in Neural
  Information Processing Systems}, vol.~34, pp.~14938--14954, 2021.

\bibitem{chung2022score}
H.~Chung and J.~C. Ye, ``Score-based diffusion models for accelerated mri,''
  {\em Medical Image Analysis}, p.~102479, 2022.

\bibitem{tiepolt2016early}
S.~Tiepolt, S.~Hesse, M.~Patt, J.~Luthardt, M.~L. Schroeter, K.-T. Hoffmann,
  D.~Weise, H.-J. Gertz, O.~Sabri, and H.~Barthel, ``Early [18f] florbetaben
  and [11c] pib pet images are a surrogate biomarker of neuronal injury in
  alzheimer’s disease,'' {\em European journal of nuclear medicine and
  molecular imaging}, vol.~43, no.~9, pp.~1700--1709, 2016.

\bibitem{hammes2017multimodal}
J.~Hammes, I.~Leuwer, G.~N. Bischof, A.~Drzezga, and T.~van Eimeren,
  ``Multimodal correlation of dynamic [18f]-av-1451 perfusion pet and neuronal
  hypometabolism in [18f]-fdg pet,'' {\em European journal of nuclear medicine
  and molecular imaging}, vol.~44, no.~13, pp.~2249--2256, 2017.

\bibitem{visser2020tau}
D.~Visser, E.~E. Wolters, S.~C. Verfaillie, E.~M. Coomans, T.~Timmers,
  H.~Tuncel, J.~Reimand, R.~Boellaard, A.~D. Windhorst, P.~Scheltens, {\em
  et~al.}, ``Tau pathology and relative cerebral blood flow are independently
  associated with cognition in alzheimer’s disease,'' {\em European journal
  of nuclear medicine and molecular imaging}, vol.~47, no.~13, pp.~3165--3175,
  2020.

\bibitem{avants2009advanced}
B.~B. Avants, N.~Tustison, G.~Song, {\em et~al.}, ``Advanced normalization
  tools (ants),'' {\em Insight j}, vol.~2, no.~365, pp.~1--35, 2009.

\bibitem{fischl2012freesurfer}
B.~Fischl, ``Freesurfer,'' {\em Neuroimage}, vol.~62, no.~2, pp.~774--781,
  2012.

\bibitem{qi2006iterative}
J.~Qi and R.~M. Leahy, ``Iterative reconstruction techniques in emission
  computed tomography,'' {\em Physics in Medicine \& Biology}, vol.~51, no.~15,
  p.~R541, 2006.

\bibitem{chung2022come}
H.~Chung, B.~Sim, and J.~C. Ye, ``Come-closer-diffuse-faster: Accelerating
  conditional diffusion models for inverse problems through stochastic
  contraction,'' in {\em Proceedings of the IEEE/CVF Conference on Computer
  Vision and Pattern Recognition}, pp.~12413--12422, 2022.

\end{thebibliography}
